\documentstyle[12pt]{article}

\begin{document}
\newcommand{\be}{\begin{equation}}
\newcommand{\ee}{\end{equation}}
\newcommand{\al}{\alpha}
\title
{
\vspace*{-20mm}
\bf
\begin{flushright}
{\large \bf TTP/98-13\\[-5mm]
MZ-TH/98-11\\[-5mm]
March 1998\\[15mm]}
\end{flushright}
\bf
Next-to-next-to-leading order vacuum polarization function of heavy quark
near threshold and sum rules for  
$b \bar b$ system.
}
\author{
  A.A.Penin\thanks{On leave from Institute for Nuclear Research,
  Moscow, Russia}\\[-5mm] 
  {\small {\em Institut f{\"u}r Theoretische Teilchenphysik
  Universit{\"a}t Karlsruhe}}\\[-5mm]
  {\small {\em D-76128 Karlsruhe, Germany}}\\
  and\\
  A.A.Pivovarov{$^*$}\\[-5mm]
  {\small {\em 
Institut f\"ur Physik, Johannes-Gutenberg-Universit\"at}}\\[-5mm]
  {\small {\em Staudinger Weg 7, D-55099 Mainz, Germany}}
}
\date{}

\maketitle

\begin{abstract}
A correlator of the vector current of a heavy quark
is computed analytically near threshold 
in the next-to-next-to-leading order in perturbative and relativistic 
expansion that 
includes $\al_s^2$, $\al_sv$ and $v^2$ corrections
in the coupling constant and velocity of the heavy quark
to the  nonrelativistic Coulomb approximation.
Based on this result, the numerical values of the $b$-quark pole mass
and the strong coupling constant are determined 
from the analysis of sum rules for
the $\Upsilon$ system. The next-to-next-to-leading corrections
are found to be of order of next-to-leading ones.\\[2mm]
\end{abstract}

\thispagestyle{empty}

\newpage

\noindent 
Insufficiency of the ordinary PT for description the near threshold
behavior of vacuum polarization function was noted long ago in the context
of Coulombic resummation in nonrelativistic QED \cite{Braun,Barb}.
Recently a considerable progress has been made in studying the 
near threshold production of heavy quark-antiquark pair within 
perturbation theory of QCD with resummation of threshold singularities.
Both perturbative and relativistic corrections have been taken into
account in the 
next-to-next-to-leading order
in the coupling constant and velocity of the heavy quark
to the leading nonrelativistic approximation based on Coulomb
potential \cite{KPP,Hoang,Mel}. 
This theoretical development provides more accurate 
description of the heavy quark vacuum polarization 
function in the threshold region necessary for
such applications as the top
quark production \cite{top} and the precise quantitative 
investigation of the $\Upsilon$
system \cite{NSVZ,Vol2}. In the latter case
higher order corrections to leading Coulomb behavior in the threshold
region
are essential both 
numerically for extracting the 
$b$-quark mass and the strong coupling constant \cite{KPP}
and qualitatively for justifying the perturbative expansion
around Coulomb solution. 
The analytical calculation  of    
the next-to-next-to-leading  order corrections 
has not 
been completed yet though some results are available\footnote{
Semi-analytical analysis of the complete next-to-next-to-leading
order corrections to the heavy quark polarization  function
near the two-particle threshold has been done 
in the context of the photon mediated t quark pair production
\cite{Hoang,Mel}.}. 
  
In this  paper we present the complete analytical expression for 
a correlator of the 
vector current of heavy
quarks near threshold 
in the next-to-next-to-leading  order
resumming  all ${\cal O}[(\al_s /v)^n\times(\al_s^2,\al_sv,v^2)]$ 
terms, with $v$ being the heavy quark velocity.
The correlator is further used 
for determination of the bottom quark pole mass $m_b$
and the strong coupling constant $\al_s$ from sum rules for
the $\Upsilon$ system.

We study the near 
threshold behavior of the polarization function  $\Pi(s)$
of the $b$-quark vector current $j_\mu=\bar b\gamma_\mu b$
\[
\left(q_\mu q_\nu-g_{\mu \nu}q^2\right)\Pi(q^2)=
i\int dxe^{iqx}\langle 0|Tj_{\mu}(x)j_{\nu}(0)|0\rangle 
\]
within the nonrelativistic expansion \cite{CasLep} which 
in the next-to-next-to-leading order reads
\be
\Pi(s)
={N_c\over 2 m_b^2}\left(
C_h(\al_s)G(0,0,k)+{4\over 3}{k^2\over m_b^2}G_C(0,0,k)\right)
\label{thmom}
\ee
with $k=\sqrt{m_b^2-s/4}$ being a natural energy variable near
threshold.
First term in brackets gives the representation for the correlator
within NRQCD with
$C_h(\al_s)$ being a
perturbative coefficient matching
correlators of relativistic and nonrelativistic vector currents. 
The coefficient 
$C_h(\al_s)$ is computable in full QCD and by now is 
known to the second order in $\al_s$ expansion
\[
C_h(\al_s)
=1-C_h^1C_F{\al_s\over \pi}+C_h^2C_F\left({\al_s\over \pi}\right)^2
\]
with 
$
C_h^1=4$ \cite{KS}
and 
\[
C_h^2=\left({39\over 4}-\zeta(3)+{4\pi^2\over 3} 
\ln{2}-{35\pi^2\over 18}\right)C_F-\left({151\over 36}+{13\over 2}
\zeta(3)+{8\pi^2\over 3} \ln{2}-{179\pi^2\over 72}\right)C_A
\]
\be
+
\left({44\over 9}-{4\pi^2\over 9}+{11\over 9}n_f\right)T_F
+2\left(\beta_0+{\pi^2\over 3}C_F+{\pi^2\over 2}C_A
\right)\ln{\left(m_b\over\mu\right)}
\label{ch}
\ee
in $\overline{\rm MS}$ renormalization scheme \cite{Hoang,Mel,ch}. 
Here the group invariants for QCD are $C_A=3$, $C_F=4/3$, $T_F=1/2$,
and $\gamma_E=0.577216\ldots$ is the Euler constant, $\zeta(z)$ is 
the Riemann $\zeta$-function, $n_f$ is the number of light flavors, and
$
\beta_0=11C_A/3-4T_Fn_f/3$.
The quantity $G({\bf x},{\bf y},k)$ is the nonrelativistic 
Green function (GF) of  the following 
Schr{\"o}dinger equation
\[
\left(-{\Delta_{\bf x}\over m_b}-{\Delta_{\bf x}^2\over 4m_b^3}
+V_C(x)+{\al_s\over 4\pi}V_1(x)+\left({\al_s\over 4\pi}\right)^2
V_2(x)\right.
\]
\be
\left.
+V_{NA}(x)+V_{BF}({\bf x},{\bf s}) 
+ {k^2\over m_b}\right)
G({\bf x},{\bf y},k)=\delta({\bf x}-{\bf y})
\label{Schr}
\ee
where $V_C(x)=-C_F\al_s/x$ is the  Coulomb
potential which is supposed to dominate the whole QCD interaction in the
energy region of interest, $x=|{\bf x}|$,
$V_{NA}(x)=-C_AC_F\al_s^2/(m_bx^2)$ is the non-Abelian potential 
of quark-antiquark interaction \cite{Gup}, $V_{BF}({\bf x},{\bf s})$ 
is the standard Breit-Fermi potential (up to the color factor $C_F$)
containing the quark 
spin operator ${\bf s}$, {\it e.g.}~\cite{Landau}.
The terms $V_i$ ($i=1,2$) represent
first and second order perturbative QCD corrections to the  Coulomb
potential \cite{Fish,Peter}
$$
V_1(x)=V_C(x)(C_0^1+C_1^1\ln(x\mu)),
$$
\be
V_2(x)=V_C(x)(C_0^2+C_1^2\ln(x\mu)+C_2^2 \ln^2(x\mu)),
\label{potcorr1}
\ee
where  
\[
C_0^1=a_1+2\beta_0\gamma_E,\qquad C_1^1=2\beta_0,
\]
\[
C_0^2=\left({\pi^2\over 3}+4\gamma_E^2\right)\beta_0^2
+2(\beta_1+2\beta_0a_1)\gamma_E+a_2,
\]
\[
C_1^2=2(\beta_1+2\beta_0a_1)+8\beta_0^2\gamma_E,
\qquad 
C_2^2=4\beta_0^2,
\]
$$
a_1={31\over 9}C_A-{20\over 9}T_Fn_f,
$$
\[
a_2= \left({4343\over 162}+6\pi^2-{\pi^4\over 4}
+{22\over3}\zeta(3)\right)C_A^2-
\left({1798\over 81} + {56\over 3}\zeta(3)\right)C_AT_Fn_f
\]
$$
-\left({55\over 3} - 16\zeta(3)\right)C_FT_Fn_f
+\left({20\over 9}T_Fn_f\right)^2,
$$
\[
\beta_1={34\over 3}C_A^2-{20\over 3}C_AT_Fn_f-4C_FT_Fn_f.
\]
The second term in eq.~(\ref{thmom}) is generated by the 
operator of dimension four in  
the  nonrelativistic expansion of the 
vector current (see, for example, \cite{Bod}). It contains 
the GF of the pure Coulomb Schr{\"o}dinger equation \cite{Schw}
at the origin 
\be
G_C(0,0,k)=-{C_F\al_sm_b^2\over 4\pi}\left({k\over C_F\al_sm_b}+
\ln\left({k\over \mu}\right)+\gamma_E+\Psi_1\left(1- {C_F\al_sm_b\over 2k}
\right)\right)
\label{G0}
\ee
where $\Psi_1(x)={\Gamma'(x)/\Gamma(x)}$
and $\Gamma(x)$ is the Euler $\Gamma$-function.

The solution to eq.~(\ref{Schr}) can be 
found within the standard nonrelativistic 
perturbation theory
around the Coulomb GF $G_C({\bf x},{\bf y},k)$.
The leading order 
corrections to the Coulomb GF at the origin
due to 
$\Delta^2$, $V_{NA}$ and  $V_{BF}$ terms are known 
analytically 
\cite{Hoang,Mel}\footnote{The term $V_{NA}$ can be fully 
accounted for the
Coulomb GF 
because the corresponding differential 
equation is exactly solvable in
standard special functions. Numerically this is not important for
applications though.}.
After including these corrections 
the approximate GF of eq.~(\ref{Schr}) at the origin 
takes the form \cite{Hoang}
\[
G(0,0,k)=-
{C_F\al_sm_b^2\over 4\pi}\left(
\left(1-{5\over 8}{k^2\over m_b^2}\right){k\over C_F\al_sm_b}+
\left(1-2{k^2\over m_b^2}\right)\right.
\]
\be
\left.\left(
\ln\left({k\over \mu}\right)+\gamma_E
+\Psi_1\left(1- 
{C_F\al_sm_b\over 2k}\right)\right)
+{11\over 16}{C_F\al_s k\over m_b}
\Psi_2\left(1-{C_F\al_sm_b\over 2k}\right)
\right)
\label{Hoangcorr}
\ee
\[
+{4\pi\over 3}{C_F\al_s\over m_b^2}
\left(1+{3\over 2}{C_A\over C_F}\right)G_C^2(0,0,k)
\]
where $\Psi_2(x)=\Psi_1'(x)$.
Note that in  ref.~\cite{Hoang} the shift of the
spectrum of  intermediate nonrelativistic Coulomb bound states
was treated exactly {\it i.e.} without expanding of 
the  energy  denominators. This accounts for
a part of the higher order corrections. We, however, consistently
work in the next-to-next-to-leading order and keep only 
the second order terms  in  eq.~(\ref{Hoangcorr}). 
Since this part of the corrections is relatively small
the difference between these two approaches is 
really negligible for the numerical analysis of the sum rules.

The correction $\Delta G_1$ to eq.~(\ref{Hoangcorr}) 
due to the first iteration
of $V_1$ term of the QCD potential 
has been found in ref.~\cite{KPP} where
the consistent analysis of sum rules for $b\bar b$ system in 
the next-to-leading order 
has been performed
$$
\Delta G_1(0,0,k)={\al_s\over 4\pi}{C_F\al_sm_b^2\over 4\pi}\left(
\sum_{m=0}^\infty F^2(m)(m+1)
\left(C_0^1+(L_k+\Psi_1(m+2))C_1^1\right)
\right.
$$
\[
-2\sum_{m=1}^\infty\sum_{n=0}^{m-1}
F(m)F(n)
{n+1\over m-n}C_1^1 
+2\sum_{m=0}^\infty F(m)
\left(C_0^1+(L_k -
2\gamma_E-\Psi_1(m+1))C_1^1\right)
\]
\be
\left.+L_kC_0^1+\left(-\gamma_E L_k+
{1\over 2}L_k^2\right)C_1^1
\right)
\label{G1}
\ee
where $L_k=\ln\left({\mu \over 2k}\right)$ and 
\[
F(m)={C_F\al_s m_b\over (m+1)2k}\left(m+1-{\displaystyle {C_F\al_s m_b\over 2k}}\right)^{-1}. 
\] 
The  correction $\Delta G_2^{(2)}$ to eq.~(\ref{Hoangcorr})
due to  $V_2$ part of the potential 
is also known \cite{KPP}
\[
\Delta G_2^{(2)}(0,0,k)=\left({\al_s\over 4\pi}\right)^2{C_F\al_sm_b^2\over
4\pi}\left( \sum_{m=0}^\infty F^2(m)
\left((m+1)\left(C_0^2+L_k C_1^2+L_k^2 C_2^2\right)
\right.\right.
\]
$$
\left.
+(m+1)\Psi_1(m+2)\left(C_1^2+2L_k
C_2^2\right)+I(m)C_2^2\right) 
$$
\be
+2\sum_{m=1}^\infty\sum_{n=0}^{m-1}
F(m)F(n)\left(-
{n+1\over m-n}\left(C_1^2 +2L_k C_2^2\right)
+J(m,n)C_2^2\right)
\label{G22}
\ee
\[
+2\sum_{m=0}^\infty F(m)
\left(C_0^2+L_k C_1^2+
(L_k^2+K(m))C_2^2-(2\gamma_E+
\Psi_1(m+1))
\left(C_1^2+2L_k C_2^2\right)\right)
\]
$$
\left.+L_kC_0^2+\left(-\gamma_E L_k+
{1\over 2}L_k^2\right)
C_1^2 
+N(k)C_2^2\right)
$$
where
\[
I(m)=(m+1)\left(\Psi^2_1(m+2)-\Psi_2(m+2)+{\pi^2\over3}-{2\over(m+1)^2}\right)
\]
$$
-2(\Psi_1(m+1)+\gamma_E),
$$
$$
J(m,n)= 2{n+1\over m-n}\left(\Psi_1(m-n)-{1\over n+1}+2\gamma_E\right)
$$
\[
+2{m+1\over m-n}(\Psi_1(m-n+1)-\Psi_1(m+1)),
\]
$$
K(m)=2(\Psi_1(m+1)+\gamma_E)^2+\Psi_2(m+1)-\Psi_1^2(m+1)+2\gamma_E^2,
$$
\[
N(k)=\left(\gamma_E+{\pi^2\over 6}\right)L_k
-\gamma_E L_k^2+{1\over 3}
L_k^3.
\]
In this paper we complete these results by computing 
the correction $\Delta G_2^{(1)}$ due to the 
second iteration of $V_1$ term which of the proper 
(next-to-next-to-leading) order according to counting
of smallness in nonrelativistic QCD with respect to $\al_s$ and $v$.
The result reads
\[
\Delta G_2^{(1)}(0,0,k)=\left({\al_s\over 4\pi}\right)^2
{(C_F\al_s)^2\over 4\pi}{m_b^3\over 2k}
\left( \sum_{m=0}^\infty H^3(m)(m+1)
\right.
\]
$$
\left(C_0^1+
\left(\Psi(m+2)+
L_k\right)C_1^1\right)^2
$$
\[
-2\sum_{m=1}^\infty\sum_{n=0}^{m-1}{n+1\over m-n}C_1^1
\left(H^2(m)H(n)\left(C_0^1+\left(\Psi(m+2)+
L_k-{1\over 2}{1\over m-n}\right)C_1^1\right)
\right.
\]
\be
\left.
+H(m)H^2(n)\left(C_0^1+\left(\Psi(n+2)+
L_k-{1\over 2}{n+1\over (m-n)(m+1)}\right)C_1^1\right)\right)
\label{G21}
\ee
\[
+2(C_1^1)^2\left(\sum_{m=2}^\infty\sum_{l=1}^{m-1}\sum_{n=0}^{l-1}
{H(m)H(n)H(l)}{n+1\over (l-n) (m-n)}\right.
\]
$$
+\sum_{m=2}^\infty\sum_{n=1}^{m-1}\sum_{l=0}^{n-1}
{H(m)H(n)H(l)}{l+1\over (n-l)(m-n)}
$$
\[
\left.\left.
+\sum_{n=2}^\infty\sum_{m=1}^{n-1}\sum_{l=0}^{m-1}
{H(m)H(n)H(l)}{(l+1)(m+1)\over (n+1)(n-l)(n-m)}\right)\right)
\]
where
$$
H(m)=\left(m+1-{\displaystyle {C_F\al_s m_b\over 2k}}\right)^{-1}. 
$$ 
We are going to describe the details of 
this rather cumbersome
calculation 
elsewhere. One remark is in order though.  
Because
ultraviolet divergences in eq.~(\ref{Hoangcorr}) 
depend on $k$ one has 
to  match the calculation of these corrections to the 
calculation of the Wilson coefficient $C_h(\al_s)$ (eq.~(\ref{ch}))
\cite{Hoang,Mel}. 
Such matching is not necessary for the calculation of $\Delta G_1$ and
$\Delta G_2^{(i)}$
terms because their divergent parts are $k$ independent.

Thus 
eqs.~(\ref{thmom},~\ref{G0}-\ref{G21}) give 
the complete analytical expressions for the vacuum polarization
function of heavy quarks near the two-particle 
threshold in the next-to-next-to-leading order\footnote{In 
refs.~\cite{Hoang,Mel} the corrections to the 
Coulomb GF due to  $V_i$ ($i=1,2$) 
terms of the potential were treated numerically for complex values 
of energy far from the real axis.}. Eqs. (\ref{G0}-\ref{G21}) look
awkward and they can be rendered into more readable form by using
$\Psi$ functions for expressing some of the sums entering the formulae.
However for direct numerical analysis of sum rules for $b\bar b$
system this form is most suitable with respect to applicability of
efficient numerical algorithms of a symbolic system.

Obtained formulae are applied to 
the analysis of the 
$\Upsilon$ system for extraction of the 
$b$-quark pole mass $m_b$ and the coupling
constant $\al_s$.
The sum rules are formulated in the literature \cite{KPP,Vol2}
and we will use the latest version \cite{KPP} with correct large $n$
behavior.
The moments ${\cal M}_n$ 
\[
{\cal M}_n= 
\left.{12\pi^2\over n!}(4m_b^2)^n{d^n\over ds^n}\Pi(s)\right|_{s=0}=
(4m_b^2)^n\int_0^\infty{R(s)ds\over s^{n+1}}
\]
of the spectral density 
$R(s)=12\pi {\rm Im}\Pi(s+i\epsilon)$ 
are compared with experimental ones 
\[
{\cal M}_n^{exp} = 
{(4m_b^2)^n\over Q_b^2}\int_0^\infty{R_b(s)ds\over s^{n+1}}
\]
under the assumption of quark-hadron duality.
The experimental moments ${\cal M}_n^{exp}$ are generated by 
the function $R_b(s)$ which is the 
normalizad cross section
$R_b(s) = \sigma(e^+e^-\rightarrow {\rm hadrons}_{\,b\bar b})/
\sigma(e^+e^-\rightarrow \mu^+\mu^-)$.
Here $Q_b=-1/3$ is the $b$-quark electric charge.

Numerical values
are obtained basically by saturating 
the experimental moments with the contribution
of
the first six $\Upsilon$ resonances (see \cite{KPP} for details). 
Their leptonic widths
$\Gamma_{k}$ and masses $M_{k}$ $(k=1\ldots 6)$
are known with good accuracy \cite{PDG}
\[
{\cal M}_n^{exp}={(4m_b^2)^n\over Q_b^2}
\left({9\pi\over \al_{QED}^2(m_b)}
\sum_{k=1}^6{\Gamma_{k}\over M_{k}^{2n+1}}
+\int_{s_0}^\infty\!{\rm d}s {R_b(s)\over s^{n+1}}\right).
\]
The rest of the spectrum beyond the resonance
region  for energies larger than
$s_0\approx (11.2~{\rm GeV})^2$ (continuum contribution)
lies far from threshold and is safely 
approximated by
the ordinary PT expression for the theoretical 
spectral density, so there $R_b(s)\approx R(s)$. 
The influence of the continuum on high moments
is almost negligible numerically 
and in any case under strict control\footnote{The expressions 
for the first few moments  of the spectral density 
are now available 
in ordinary perturbation theory 
with $\al_s^2$ accuracy \cite{Chet}, however,  they cannot
be used in theoretical formulas for sum rules directly
because the spectrum is well known
experimentally only for energies close
to threshold due to existence of sharp resonances while the
contribution of the continuum to these low moments is large in
comparison with the resonance contribution.}.
Electromagnetic coupling constant is renormalized to the 
energy of order 
of $m_b$ with the result $\al_{QED}^2(m_b)=1.07 \al^2$ \cite{PDG}.

We work with moments for $10<n<20$ 
that simultaneously guarantees the smallness of 
both the continuum contribution
and the nonperturbative power corrections due to the
gluonic condensate \cite{Vol2}.
The first one
is not well known experimentally
and has to be suppressed to make results 
independent of $s_0$. The second one should be small 
because the value of gluonic condensate (and higher order condensates)
is not known well numerically.
The normalization point $\mu = m_b$ is used throughout 
the computation\footnote{
We work strictly in the next-to-next-to-leading order
approximation 
and, therefore, use the same normalization point 
for soft and hard 
corrections in contrast to \cite{Hoang,Mel} where 
different normalization points were chosen for these two parts.}. 
At the scale  $\mu =m_b$ both the hard and soft gluon
corrections are of the same order of magnitude. For a lower scale the
hard corrections become large while for higher scale
the same is true for the soft corrections \cite{KPP}.
We found that at $\mu \sim m_b$ the $\mu$ dependence
of the results is minimal which is a solid indication 
that 
at this point the 
higher order corrections are also small.

The result of the fit is 
\[
\al_s(m_b)= 0.22 \pm 0.02, \quad {\rm or}\quad
\al_s(M_Z)=0.118 \pm 0.006.
\]
The sum rules are much more sensitive to the
$b$-quark mass than to the strong coupling constant so it is instructive
to fix $\al_s(M_Z)=0.118$ to  the ``world average'' value \cite{PDG}
and then to extract $m_b(n)$. In this way we obtain
the following estimate for the mean value over the considered range of
$n$
\[
m_b=4.78~{\rm GeV}.
\]
This value is in a good agreement with the results of the 
first order analysis \cite{KPP}
where at 
$\al_s(M_Z)=0.118$ we obtained 
\[
m_b=4.75~{\rm GeV}. 
\]
Note that the optimization
procedure \cite{PenPiv} 
was used to improve convergence of perturbation 
theory in the previous 
analysis \cite{KPP}. As we see this procedure turns out to be 
a  powerful tool to  estimate  the higher order contributions.
For comparison, the leading order result is $m_b=4.70~{\rm GeV}$
and in the next-to-leading approximation without optimization one
gets  $m_b=4.72~{\rm GeV}$.

Main uncertainties of numerical values for 
considered parameters stem from the same sources that were identified 
in ref. \cite{KPP}.
The  error coming from  $n$ distribution for the mass at fixed value
of the coupling constant 
is about $\pm 0.5\%$ for $10<n<20$. The $\mu$
dependence for $\mu=m_b\pm 1~{\rm GeV}$
(where this dependence is minimal) introduces another $\pm 0.5\%$
of uncertainty.
Thus our final estimate of the bottom quark pole mass is 
\[
m_b=4.78\pm 0.04~{\rm GeV}. 
\] 
Note that the uncertainty originated from the $n$ and  $\mu$
dependence is not reduced
in comparison with the next-to-leading order.
This means that the contribution of the higher order
corrections which has to cancel  $n$ and $\mu$
dependence of the results is still important.
Let us emphasize 
that the convergence of the perturbation theory
for  the vacuum polarization function of heavy quark
near threshold is not fast.
We  have found   the next-to-next-to-leading order
corrections  to be of order of the next-to-leading ones.
Furthermore, in the case of $b$-quark the corrections  due to the 
perturbative modification of the Coulomb 
instantaneous potential ({\it i.e.} related to $\Delta G_1$ and 
$\Delta G_2^{(i)}$ terms) dominate
the total correction in the next-to-leading and 
next-to-next-to-leading orders.
Inclusion of these  corrections is quite important for 
consistent analysis of sum rules for the $\Upsilon$
system.

To conclude we have constructed an expression for the vacuum polarization
function of the vector current of a heavy quark
near threshold. It is completely analytic  
in the next-to-next-to-leading order in perturbative and relativistic 
expansion up to  
$\al_s^2$, $\al_sv$ and $v^2$ corrections. 
The polarization
function was used for determination of the $b$-quark pole mass and the
coupling constant from sum rules for the $\Upsilon$ system
that are saturated by contributions near threshold.
In fact, there is no much hope for
improving our results: next order approximation seems to be too
complicated for analytical treatment within the regular perturbation
theory for NRQCD.
The analysis showed a remarkable stability with respect to the
next-to-leading one supplied with an optimization procedure in
a variational spirit. Having in mind the considerable technical
difficulty of computing next approximation and recognizing the
necessity of improving the theoretical predictions in view of 
new high quality  experimental data we think that the next step 
in the near future will be connected with optimization of the 
present approximation. 

\vspace{3mm}
\noindent
{\large \bf Acknowledgements}\\[2mm]
We thank J.H.K{\"u}hn
for support, encouragement, and discussions.
A.A.Penin greatfully acknowledges discussions
with K.Melnikov.
This work is partially supported 
by Volkswagen Foundation under contract
No.~I/73611. A.A.Pivo\-varov is  
supported in part by
the Russian Fund for Basic Research under contracts Nos.~96-01-01860
and 97-02-17065. 
The work of A.A.Penin is supported in part  by
the Russian Fund for Basic Research under contract
97-02-17065.


\begin{thebibliography}{99}
\bibitem{Braun} M.A.Braun, ZhETP Lett. {\bf 27}(1968)652.
\bibitem{Barb}  R.Barbieri, P.Christillin and E.Remiddi, 
Phys.Rev. {\bf A8}(1973)2266.
\bibitem{KPP}  J.H.K{\"u}hn, A.A.Penin and A.A.Pivovarov, Preprint 
               {\bf TTP-98-01}, \\ hep-ph/9801356.
\bibitem{Hoang} A.H.Hoang and T.Teubner, Preprint {\bf UCSD/PTH 98-01},
                hep-ph/9801397.  
\bibitem{Mel}    K.Melnikov  and A.Yelkhovsky, Preprint 
                 {\bf TTP-98-10}, hep-ph/9802379. 
\bibitem{top}  V.S.Fadin and V.A.Khoze, Pis'ma Zh.Eksp.Teor.Fiz 
               {\bf 46}(1987)417; Yad.Fiz. {\bf 48}(1988)487;\\
               W.Kwong,  Phys.Rev. {\bf D43}(1991)1488;\\
               M.J.Strassler and M.E.Peskin,  Phys.Rev. {\bf D43}(1991)1500;\\
               M.Jezabek,  J.H.K{\"u}hn and T.Teubner, Z.Phys. 
               {\bf C56}(1992)653;\\
               Y.Sumino, K.Fujii, K.Hagivara, H. Murayama and C.-K.Ng,
               Phys.Rev. \\{\bf D47}(1993)56;\\
               K.Fujii, T.Matsui and Y.Sumino,  Phys.Rev. {\bf
               D50}(1994)4341.

\bibitem{NSVZ} V.A.Novikov {\it et al.}, Phys.Rev.Lett. {\bf 38}(1977)626;\\
               V.A.Novikov {\it et al.}, Phys.Rep. {\bf C41}(1978)1;\\
               M.B.Voloshin, Yad.Fiz. {\bf 36}(1982)247;\\
               M.B.Voloshin and Yu.M.Zaitsev, Usp.Fiz.Nauk
               {\bf 152}(1987)361.

\bibitem{Vol2} M.Voloshin, Int.J.Mod.Phys. {\bf A10}(1995)2865.


\bibitem{CasLep} W.E.Caswell and G.E.Lepage, Phys.Lett.   {\bf B167}(1986)437.

\bibitem{KS}    G.K{\"a}llen and A.Sarby, 
                K.Dan.Vidensk.Selsk.Mat.-Fis.Medd. {\bf 29}(1955), N17, 1.

\bibitem{ch}     A.H.Hoang, Phys.Rev. {\bf D56}(1997)7276;\\
                 A.H.Hoang, J.H.K{\"u}hn and T.Teubner,  
                 Nucl.Phys. {\bf B452}(1995)173;\\
                 A.Czarnecky and K.Melnikov, Preprint 
                 {\bf TTP-97-54}, hep-ph/9712222;\\
                 M.Beneke, A.Signer and V.A.Smirnov,  Preprint 
                 {\bf CERN-TH-97-353}, \\ hep-ph/9712302. 
             

\bibitem{Gup}  S.N.Gupta and S.F.Radford,  Phys.Rev. {\bf D24}(1981)2309;
               Phys.Rev. \\{\bf D25}(1982)3430 (Erratum);\\
               S.N.Gupta, S.F.Radford and W.W.Repko, 
                Phys.Rev. {\bf D26}(1982)3305. 

\bibitem{Landau} L.D.Landau and E.M.Lifshitz, Relativistic Quantum
                 Theory, Part 1 (Pergamon, Oxford, 1974).

\bibitem{Fish}   W.Fisher, Nucl.Phys {\bf B129}(1977)157;\\
                 A.Billoire, Phys.Lett. {\bf B92}(1980)343. 

\bibitem{Peter}M.Peter,  Phys.Rev.Lett. {\bf 78}(1997)602;
               Preprint {\bf TTP-97-03}, hep-ph/9702245.

\bibitem{Bod}  G.T.Bodwin, E.Braaten and G.P.Lepage,   
               Phys.Rev. {\bf D51}(1995)1125. 

\bibitem{Schw} J.Schwinger, J.Math.Phys. {\bf 5}(1964)1606. 


\bibitem{PDG} Particle Data Groop, Phys.Rev. {\bf D54}(1996)1.

\bibitem{Chet} K.G.Chetyrkin, J.H.K\"uhn and  M.Steinhauser,
               Phys.Lett. {\bf B371}(1996)93;\\
               Nucl.Phys. {\bf B482}(1996)213.


\bibitem{PenPiv}  A.A.Penin and  A.A.Pivovarov, Phys.Lett.
                  {\bf B367}(1996)342.

\end{thebibliography}
\end{document}